\documentclass[12pt]{article}
\usepackage{a4}
\usepackage{epsf}

\begin{document}

\begin{center}
\Large{Local Impurity Phase Pinning and Pinning Force}\\
\Large{in Charge Density Waves}\\[1cm]

\large{A. Kobelkov, F. Gleisberg}\\
\large{and W. Wonneberger}\\[1cm]

\large{Department of Physics}\\
\large{University of Ulm}\\[0.5cm]
\large{D 89069 Ulm/Germany}
\end{center}

\noindent
\begin{abstract}
Starting from the static Fukuyama--Lee--Rice equation 
for a three--di\-men\-sio\-nal
incommensurate charge density wave (CDW) in quasi one--dimensional conductors
a solvable model for local phase pinning by impurities is defined and studied.
We find that
average CDW energy and average pinning force show critical behaviour with
respect to the pinning parameter $h$. Specifically the pinning force exhibits
a threshold at $h=1$ with exponent $\beta=2$. Our model examplifies a general
concept of local impurity pinning in which the force exerted by the impurity
on the periodic CDW structure becomes multivalued and metastable states appear
beyond a threshold. It is found that local impurity pinning becomes less
effective at low temperatures and may eventually cease completely.
These results are independent of spatial dimensionality as expected for local
impurity pinning. Comparison with Larkin's model is also made.
\end{abstract}


\section{Introduction}

In charge density waves (CDW) which appear below the Peierls
transition temperature in quasi
one--dimensional metals \cite{G88}, pinning of the order parameter phase at
point defects is an important effect. Phase pinning results from
the electrostatic coupling between the spatially periodic charge modulation
in the CDW and the electric potential of an impurity.
The phase $\varphi$ determines the position of the CDW with respect
to the host lattice. The energy and force contributions of an individual
impurity become periodic functions of $\varphi $.
Distortions of the phase near the impurity produce a positive elastic
energy. By properly
adjusting the phase pattern near the impurity a net pinning force
can be possible.
Phase pinning in CDW is not fully understood at present.
Well known is the weak or collective pinning
limit. In this case each impurity only slightly distorts the local
phase, however, many impurities act coherently in a macroscopic metastable
Lee--Rice domain \cite{FL78,LR79} and produce a small pinning force.
This mechanism has its analogue in pinning
of flux lines in type II superconductors \cite{LO73,LO79}. The opposite 
limit of strong pinning is obtained when the pinning strength of
a single impurity is so large that the elastic energy of
local phase deformations is neglegible \cite{FL78}. Intermediate concepts have
also been explored \cite{A86,LLS87,TGL88}.

In this paper we point out a number of special features for local phase
pinning by impurities.
Local phase pinning describes pinning effects linear in the concentration
of impurities. We study a
three--dimensional incommensurate CDW within a solvable model. We find that
the averaged CDW energy shows a weak singularity as function
of the pinning parameter $h$ at $h = 1$ characterized by an exponent $b$
which is calculated. The formal limit of strong pinning is reached
only for very large $h$.
The pinning force has a threshold at $h=1$, i.e., it vanishes for
$h \le 1$ and then starts with a ``critical'' exponent similar to a
phase transition. Our model realizes a general concept of local impurity
pinning when the
force exerted by the impurity on the periodic structure (CDW in our case)
becomes multivalued and metastable states appear \cite{LO79}. In a single
chain model a similar problem
has been considered recently in \cite{L94,OBLM96} neglecting the screening.

Larkin's model \cite{L94} has been extended in \cite{LB95}
to include dynamics. It has meanwhile evolved into a complete concept
of CDW and SDW dynamics \cite{B96,BL97}. Here, we restrict
ourselves to the static properties of a complementary model.

The model is introduced in Sec. 2 . It is exactly solved for average energy
(Sec. 3) and average pinning force (Sec. 4) in the screened limit when an
abundance of quasi--particles eliminates Coulomb forces. The latter act
between the local charges that are produced by phase deformations in CDW.
In Sec. 5 we discuss descreening in semiconducting CDW.
We show that in lowering the temperature local impurity pinning
becomes less and less effective and may even cease completely. Finally
Sec. 6 gives a detailed account of Larkin's model \cite{L94} and its
relation to our results.

\section{Pinning Model}

Following \cite{LR79} we write the phase dependent part of the static energy
density of incommensurate  CDW plus one impurity at position ${\bf x}_i$
as

\begin{eqnarray}\label{1}
{\cal{H}} = \frac{1}{2}\,K \left \{ \left( \frac{\partial \varphi }{\partial
z} \right) ^2 + \frac{v ^2_{t1}}{v ^2_F}
\left( \frac{\partial \varphi }{\partial x}\right) ^2 + \frac{v_{t2}^2}
{v_F^2} \left( \frac{\partial \varphi }{\partial y}\right) ^2\right\}
+ V_0 \delta ( {\bf x} - {\bf x}_i) \cos ( {\bf Q x}_i + \varphi({\bf x}) ).
\end{eqnarray}

\noindent
The stiffness constant in the fully screened limit and for a rectangular
lattice is

\begin{eqnarray}\label{2}
K = \frac{\hbar v _F}{2\pi a_{t1}a_{t2}},
\end{eqnarray}

\noindent
where $a_{t_\nu}$ are the interchain distances of the chains which
run along the z--direction. The Fermi velocities for the
anisotropic metal are denoted $v_F$, $v _{t1}$ and
$v _{t2}$, respectively. In terms of the Fourier component
$V({\bf Q)}$ of the short range impurity potential (${\bf Q}$
denotes the CDW nesting vector) and in terms of the half gap $\Delta $,
the electron density $n_0, $ Fermi
energy $\epsilon _F$, and electron--phonon coupling constant $\lambda
_Q$ the pinning amplitude $V_0$ is given by
$V_0 = n_0 V(Q)\Delta/(4\lambda _Q\epsilon _F)$ \cite{AK82}.
$V_0$ has dimension of energy. We neglect dislocation lines in the CDW
lattice and consider the phase to be unique over the crystal.

\noindent
As in \cite{LR79} we perform an isotropy scaling in the static CDW energy
$E=\int d^3x {\cal{H}}$ according to $z'=z$,
$x'=x v _F / v _{t1}$, and $y'=y v _F / v _{t2}$
and use the old names. Then ${\cal{H}}$ becomes

\begin{eqnarray}\label{3}
{\cal{H}}= \frac{1}{2} K ( \nabla \varphi)^2
+ V_0 \delta({\bf x}- {\bf x}_i) \cos({\bf Q x} + \varphi),
\end{eqnarray}

\noindent
with $K = \hbar v_{t1}v_{t2}/(2\pi a_{t1}a_{t2}v _F)$,
while $V_0$ remains unchanged. The formal solution of the Poisson equation
following from (\ref{3}) gives divergent energy in three dimensions.
We introduce a cutoff $\xi$ which is of the order of the amplitude
coherence length $\hbar v_F/\Delta$ into our model by
setting $\varphi (|{\bf x}- {\bf x}_i | \le \xi/2) \equiv \varphi _i$.
Thus the phase in  our quasi--isotropic setting
is constant inside a sphere of diameter $\xi$ centered at
the impurity. $\xi$ is a constant parameter of the model.
This case corresponds to the following modified inhomogenity
in the Poisson equation associated to (\ref{3}):

\begin{eqnarray}\label{4}
-V_0 \delta(|{\bf x}-{\bf x}_i|-\xi/2) \frac{\sin({\bf Q}{\bf x}_i +
\varphi_i)}{\pi \xi^2}.
\end{eqnarray}

\noindent
For convenience we fix the phase at infinity
by requiring $\varphi _\infty = 0$. A phase $\varphi _\infty \neq 0$ can
be trivially transformed away from all our equations, especially from
the energies to be calculated. The latter, therefore, do not directly
reflect pinning. Phase pinning, however, shows up in the pinning force
\cite{LO79,OBLM96}.

\noindent
The solution of the modified Poisson equation for the phase is

\begin{eqnarray}\label{6}
\varphi (|{\bf x}-{\bf x}_i| \ge \xi/2) = \frac{\xi}{2} \,
\frac{\varphi _i}{|{\bf x}- {\bf x}_i|}.
\end{eqnarray}

\noindent
In (\ref{3}) ${\bf Q x}_i$ acts as a random phase uniformely distributed in
($-\pi,\pi $) and we will denote it as $\Gamma $. By averaging the total energy
over $\Gamma$ we obtain the CDW energy per pin of a random ensemble of
local pins.

\noindent
The total CDW energy associated with the solution $\varphi  ({\bf x})$
becomes $E = \pi K \xi \varphi_i^2 + V_0 \cos(\Gamma+\varphi_i)$.
With the pinning parameter

\begin{eqnarray}\label{7}
h \equiv \frac{V_0}{2 \pi K \xi}
\end{eqnarray}

\noindent
the energy can be expressed as

\begin{eqnarray}\label{8}
E = V_0 \left(\frac{\varphi_i^2}{2 h} + \cos (\Gamma  + \varphi _i)\right).
\end{eqnarray}

\noindent
Finally the phase $\varphi_i$ follows from the requirement $\partial E/
\partial \varphi_i =0$ or

\begin{eqnarray}\label{9}
\varphi_i =  h \sin(\Gamma  + \varphi _i).
\end{eqnarray}

\noindent
This is an implicit equation for $\varphi_i$ in terms of $\Gamma$.
Equation (\ref{9}) also appears in the mean field model of Fisher
\cite{FIS85} for the multivalued static solutions at $h>1$ in the
presence of an external field.

\noindent
Knowing $\varphi _i (\Gamma )$ the reduced averaged CDW
energy $e(h)$ for example can be expressed as

\begin{eqnarray}\label{10}
e(h) \equiv <E> / V_0 = \frac{1}{2\pi } \int^\pi  _{-\pi }\, d\Gamma
\{ \frac{h}{2} \sin ^2
(\Gamma  + \varphi _i (\Gamma ))
+ \cos (\Gamma	+ \varphi _i (\Gamma)) \}.
\end{eqnarray}

\noindent
It is noted that (\ref{9}) can also be obtained without considering
$\varphi_i$ as a variational parameter by starting from the inhomogeneous
phase equation.

\noindent
We summarize the principal features and assumptions of our model:

\begin{enumerate}
\item Three--dimensional Fukuyama--Lee--Rice model with one
      randomly placed impurity
\item Averaging over the impurity position amounts to an ensemble
      average over many local, i.e., independent pins. Weak pinning
      is excluded.
\item Neglect of quantum tunneling and thermal fluctuations. For
      a discussion of the corresponding conditions see \cite{L94}.
      It is also noted that $V_0 \gg k_B T$ holds for strong pins.
\end{enumerate}

\noindent
The following two sections derive exact results from this model.

\section{Average CDW Energy}

We begin by investigating the average CDW energy in the equilibrium state.
This also serves as an introduction to our method of solution.

Our model is nontrivial because of equation (\ref{9}) and its variety of
solutions.

\noindent
For $h<1$ there exists one and only one
solution $\varphi _i$ for every $\Gamma $. Writing (\ref{9}) as

\begin{eqnarray}\label{11}
w=\Gamma +h \sin w;\quad w \equiv \varphi_i + \Gamma,
\end{eqnarray}

\noindent
the solution $w(\Gamma)$ maps the interval
$-\pi \le \Gamma < \pi$ one to one onto $-\pi \le w < \pi$.
It is then easy to calculate $e(h \le 1)$.
Transforming the integration over $\Gamma$ in (\ref{10}) to an
integration over $w$ using $d \Gamma/dw  = 1 - h \cos w$ gives immediately

\begin{eqnarray}\label{12}
e(h \le 1) = \frac{1}{2\pi } \int^\pi  _{-\pi } \, dw  (1-h
\cos w ) \left[\frac{h}{2} \sin ^2 w  + \cos w	\right] = -\frac{h}{4}.
\end{eqnarray}

\noindent
This argument fails for $h>1$ when more than one solution of (\ref{11})
exists. For $h > 1$ at least three solutions solutions exist.
For $h=3$ the situation is depicted in Fig. 1.
In general these solutions give different energies.
The physical relevant solution is the one with the lowest energy.
Choosing this solution for each $\Gamma$ defines the integration path.

\noindent
Formally the average energy above the threshold can be expressed as

\begin{eqnarray}\label{13}
e(h) = \frac{1}{2\pi } \sum_\nu  \left[\sin w - \frac{h}{4} w
- \frac{3h}{8} \sin 2w - \frac{h^2}{6} \sin ^3 w \right]^{w
^{(\nu)}_a}_{w ^{(\nu)}_i}.
\end{eqnarray}

\noindent
$(w ^{(\nu)}_i, w ^{(\nu)}_a)$ are the endpoints of appropriate
integration intervals along the $w$--axis giving minimum energy.
The endpoints of the path are $w_i^{(1)} = -\pi$, $w_a^{(1)} = -w_A$ and
$w_i^{(2)} = w_A$, $w_a^{(2)} = \pi$ where $w_a=\sin(w_A)$, i.e., there is
a jump from $-w_A$ to $w_A$ at $\Gamma=0$.

\noindent
The value $h=1$ is a singular point in the following sense: The
dependence of $e(h)+h/4$ on $h$ changes taking on locally the form of a
power law with exponent $b$. Eq. (\ref{13}) can be used to calculate
the  exponent $b$ defined by

\begin{eqnarray}\label{14}
e(h) +\frac{h}{4} \sim	\epsilon ^{b}, \quad \epsilon \equiv h - 1 \ll 1.
\end{eqnarray}

\noindent
The value of $w_A$ near $h=1$ can be found perturbatively.
To lowest order with respect to $\epsilon$ it is $w _A = \sqrt{6\epsilon}$.
Eq. (\ref{13}) then gives

\begin{eqnarray}\label{16}
e(h) +\frac{h}{4} =  \frac{8}{35 \pi} \sqrt{6}\,\, \epsilon ^{7/2},
\end{eqnarray}

\noindent
i.e., the  exponent is $b =7 /2$.
To complete the analytical investigations, the behaviour for $h \gg
1$ is studied. In this limit the value of $w_A$ is about $\pm \pi(1-1/h)$
which gives

\begin{eqnarray}\label{17}
e(h) = - 1 + \left( \frac{\pi ^2}{6}- 1 \right) / h, \quad h \gg 1.
\end{eqnarray}

\noindent
The usual strong pinning limit $<E>_{sp} = - V_0$
is thus asymptotically approached albeit at a rather slow rate. It
is also noted that for $ |\arccos(1/h) - \sqrt{h^2-1}| > \pi$
additional solutions exist which lead to more and more metastable
states.

\section{Average Pinning Force}

Physically  more important than the average energy $e(h)$ is the
average pinning force the CDW experiences when it is moved
adiabatically. This force is

\begin{eqnarray}\label{18}
F(h) = - \frac{Q}{2\pi} \int_0^{2\pi}d\Gamma\,\frac{dE(w(\Gamma),\Gamma)}
{d\Gamma} \equiv  Q\, V_0\, f(h).
\end{eqnarray}

\noindent
The pinning force $F(h)$ vanishes for $h \le 1$ since $dE/d\Gamma$
then is a single valued and periodic function over the interval $-\pi \le
\Gamma < \pi$. Physically it means that the pinning forces from different
local impurities cancel \cite{LO79}.

Above threshold transitions between branches of different energy occur.
Using the energy profile

\begin{eqnarray}\label{18a}
e(w,\Gamma) \equiv \frac{E}{V_0}=\frac{(w-\Gamma)^2}{2h} +\cos w
\end{eqnarray}

\noindent
and

\begin{eqnarray}\label{18b}
\frac{de(w(\Gamma),\Gamma)}{d\Gamma}=-\frac{w-\Gamma}{h}=\sin w,
\end{eqnarray}

\noindent
the reduced force $f(h)$ can be treated in anology to the energy.
The result is

\begin{eqnarray}\label{18c}
f(h) = \frac{1}{2\pi } \left[\frac{h}{4} (\cos 2w_B - \cos 2w_c)
- (\cos w_B - \cos w_c) \right].
\end{eqnarray}

\noindent
Here $w_B$ is the point of the energy minimum $e(w_B,\Gamma_c)$
into which a
a transition occurs from the critical metastable state characterized by a
horizontal inflection point in the energy profile at $(w_c,\Gamma_c)$.
The r.h.s. of (\ref{18c}) is just the difference in energy of these two
states devided by $2\pi$.

It is possible to give a more geometrical description of this process
and the resulting force: Above threshold the force from shifts
of the CDW by less than $x=\Gamma_c/Q$ where $\Gamma_c$ is given by
$\Gamma_c=\sqrt{h^2-1} - w_c$ with $w_c=\arccos(1/h)$) is elastic.
Larger shifts populate metastable states.

\noindent
For $\Gamma_c < \pi$ the averaging path is from
$-\pi$ to $-w_c$ and then from $w_B$ to $ \pi $ as shown in Fig. 1.
It runs over the bifurcation point ($\Gamma_c ,-w_c)$ and
jumps to the upper branch starting at $w_B=\Gamma_c + h\sin(w_B)$.
This is clearly seen in the energy profile $e(w,\Gamma).$

\noindent
Near $h=1$ one can use the perturbative results $w _c = \sqrt{2\epsilon }$,
$\, w _A = 2w_c$ to find

\begin{eqnarray}\label{19}
F(h) =	\frac{9}{4 \pi}\,Q\,V_0\, \epsilon^2, \quad \epsilon
\ll 1.
\end{eqnarray}

\noindent
Thus pinning sets in at the threshold $h=1$ with exponent $\beta = 2$.

For larger $h$ the situation is more complicated since
more metastable states exist. The averaging path now extends over $n$
periods in order to sample all metastable states.
The jump is from $(\Gamma_c,-w_c)$ to the point $(\Gamma_c, w_B)$ which
gives the lowest energy. $w_B$ is that value on the curve $\Gamma + h\sin(w)$
that lies nearest to $(2n-1)\pi$ where $E(w)$ has its absolute minimum.
The energy difference is $E(-w_c)- E(w_B)$ with $E(\pm w_c)=V_0(h/2+1/(2h))$.
Whenever $\Gamma_c$ crosses a value $2\pi n$ ($n=1,2,3,...$), i.e., when
$h=h_n$ where $h_n$ is found from $\sqrt{h^2_n-1}-\arccos(1/h_n)=2n \pi$
(approximate solution $h_n=2n \pi+\pi/2$)
the winding number changes from $n$ to $n+1$ and $f(h)$ is reduced
correspondingly. This leads to a saw--toothed
structure of $f(h_{n-1}  < h \le h_n) = (h/2 +1/(2h) -E(w_B))/(2 \pi n)$
($h_0 \equiv 1$).
Thus $f(h)$ oscillates around the asymptotic value $1/2 $ with approximate
period $2 \pi$ and decreasing peak to peak amplitude $\delta f(h_n)=
f(h_n)/(n+1)$. Fig. 2 displays the reduced pinning force $f(h)$. It is
noted that the unscaled force (\ref{18}) contains the pinning amplitude $V_0$
and does not saturate for $h \rightarrow \infty$ as in Larkin's model
\cite{L94}.

The threshold behaviour of the pinning force is also found in the
one--dimensional pinning model in \cite{L94,OBLM96} and is a general
feature of local impurity phase pinning \cite{LO79}.

\subsection{Pinning Parameters}

The relation of the pinning parameter $h$ to the usual measure of
pinning strength
will now be discussed. According to (\ref{8}) and taking the isotropy
scaling into account, $h$ is given by:

\begin{eqnarray}\label{20}
h = V_0 \frac{a_{t1}a_{t2}v _F}{v _{t1}v _{t2} \hbar \xi }.
\end{eqnarray}

\noindent
The standard measure \cite{MT87} for impurity energy to elastic energy in
an anisotropic three--dimensional CDW continuum is

\begin{eqnarray}\label{21}
\epsilon _i = 2\pi  V_0 a_{t1}a_{t2} c_3^{1/3} \left( \frac{v
_F^2}{v _{t1}v _{t2}}\right)^{2/3} \, \frac{1}{\hbar
v _F},
\end{eqnarray}

\noindent
where $c_3$ is the impurity concentration. The effective distance between
pins is $\ell = (c_3 v_{t1} v_{t2}/v _F^2)^{-1/3}$.
Thus $\epsilon_i$ is related to $h$ by

\begin{equation}\label{22}
\epsilon _i = 2 \pi \xi \,h /\ell.
\end{equation}

\noindent
The applicability of the theory requires $\xi$ to be smaller than $\ell$.
Using the values $v_F=5 \cdot 10^7 cm s^{-1}\approx 10 v_t$, $\Delta=
1000 k_B T$, and an impurity concentration of one ppm the ratio $\xi/\ell$
is estimated as $0.01$
One can, nevertheless, conclude that the strong pinning limit in our model,
$h \gg 1$, also requires large $\epsilon _i $. Weak pinning occurs
for $\epsilon_i < 1$. The region  $0<h \le 1$ where local pinning is absent
thus is concealed by weak pinning which is ubiquitous in less than
four dimensions.

\section{Descreening}

So far we have considered the case of full screening: There are enough
thermally excited quasiparticles (or normal carriers as in $\rm{NbSe_3}$)
to completely screen out the Coulomb
forces between charge fluctuations associated with phase deformations.
It is known that descreening stiffens the CDW.
This stiffening of the CDW leads to a corresponding increase of the
phason velocity which has been observed by neutron scattering \cite{HPS92}
and explained in \cite{VM92} as a descreening effect.
In a simple approximation involving only the condensate fraction $N<1$
one can define an effective stiffness constant

\begin{eqnarray}\label{24}
K_{eff} = \frac{K}{1-N},
\end{eqnarray}

\noindent
to take care of descreening within
the elastic CDW model \cite{K80,WT87,AV89,VM93,LB95,AW96,A97}. Following
\cite{A97} this is a reasonable approximation for $1-N > \zeta \equiv
\epsilon_t \hbar v_F/(8 e_0^2)$ ($\epsilon_t$: static transverse dielectric
constant). In the opposite limit Coulomb interactions require a different
approach. The modified stiffness constant changes the pinning parameter $h$
according to

\begin{eqnarray}\label{26}
h \rightarrow h \, \sqrt{1-N} \equiv h_{eff},
\end{eqnarray}

\noindent
since a scaling in chain direction becomes necessary to maintain quasi
isotropy. Our earlier formulae hold with this replacement. When
$h_{eff}$
becomes less than unity in decreasing the temperature local impurity
phase pinning stops. The singularity in the local pinning force at
$h_{eff}=1$ is likely to be masked by thermal fluctuations and by weak
pinning.

The fact that local pinning centers require a minimum strength to act
as strong pins is possibly the reason for the following observation in
\cite{MDMAT92}:
The CDW in $\rm{Ti}$ doped $\rm{NbSe_3}$ which is fully screened due to a
partially gapped Fermi surface exhibits weak instead of the expected strong
pinning.

\section{Relation to Larkins's model}

\noindent
We want to point out the similarities and differences in our approach
to Larkin's model \cite{L94} and fill in some additional information
about it.

Larkin uses a single chain model \cite{B90} with one impurity at $ x= x_i$.
It is defined by the energy functional:

\begin{eqnarray}\label{28}
E = \frac{E_s}{8} \int dx \left[ 1 - \cos \varphi (x) + \frac{1}{2} \left(
\frac{d\varphi (x)}{dx} \right )^2 \right]
+ V_0 [1 - \cos (\Gamma + \varphi  (x_i) ) ].
\end{eqnarray}

\noindent
In (\ref{28}) the length $x$ is scaled to $\hbar v_F /
(2t_\perp)$. The interchain coupling energy is denoted by
$ t_\perp$ and $E_s = 8 t_\perp /\pi $ is the energy of a $2\pi $
phase soliton \cite{MBKT76} ($E_s$ is called $w$ in \cite{L94}).

\noindent
>From the general theory in \cite{A97} which describes the CDW as a system
of coupled chains Larkin's model follows under three assumptions:

\begin{enumerate}
\item Phases on neighbouring chains are set to zero which turns the
interchain coupling into the Sine--Gordon type self interaction $1-\cos
\varphi $ in (\ref{28}). Thus phases on different chains are independent.
In contrast the phase varies slowly across the chains in our model, a
case more appropriate to a screened situation.

\item The low temperature or descreened limit is understood when
the quasi--particle
fraction $1 - N$ is smaller than the Coulomb coupling constant $\zeta$.

\item The value $\zeta $ of the Coulomb coupling constant mentioned in Sec.
5 is fixed at 1/8. In reality a smaller value holds because of
$v_F = O (10^7 $cm$  s^{-1})$.
\end{enumerate}

\noindent
The model (\ref{28}) has uncharged dipole solutions for which the phase
$\varphi_i = \varphi(x_i)$ at the impurity obeys the matching
condition $(h \equiv 4 V_0 /E_s)$

\begin{eqnarray}\label{30}
-2 \sin \frac{\varphi_i}{2} = h \sin w.
\end{eqnarray}

\noindent
This equation replaces our equation (\ref{11}). For any solution
$\varphi^{(1)}_i$ it has another--usually inequivalent--solution
$\varphi ^{(2)}_i$ with $\varphi
^{(2)}_i = 2\pi  - \varphi ^{(1)}_i$. The elastic energies are $E_s
[1 - \cos( \varphi ^{(1)}_i /2)]$ and $E_s[1 - \cos(\varphi ^{(2)}_i /2)]
=E_s[1 + \cos(\varphi ^{(1)}_i /2)]$.
We consider here the domains $0 \le w < 2\pi$ and $0 \le \varphi_i <
2\pi $. The model is thus characterized by one metastable state.

The mechanism of pinning force generation for $h <2$ is precisely the same
as discussed in Sec. 4: Transition from a metastable state
with relative energy $\Delta E$ which becomes a horizontal inflection point
in the energy
profile $E(w, \Gamma )$ for $\Gamma =\Gamma _c$ down to the ground state.
The average pinning force is then:

\begin{eqnarray}\label{32}
 F = \frac{Q \Delta  E}{2 \pi}.
\end{eqnarray}

\noindent
In our model this mechanism prevails for all $h >1$ and more and more
metastable states appear for increasing $h$. In
Larkin's model no horizontal inflections points exist for $h >2$.
Instead two $2\pi$--solitons are created when $\varphi_i$ changes by
$2\pi$ in a corresponding change of $\Gamma$.
This leads to $ F (h>2) = QE_s /\pi $ in \cite{L94} while
the average value of $F(h)$ increases linearily with $h$ in our model
as implied by the strong pinning concept.

It is possible to study the special case $h=2$ analytically because the
exact solution of (\ref{30}) -- expressed as $w=w(\Gamma )$ -- is available:

$$ w= \frac{\Gamma}{3} + \frac{4\pi }{3} n,\quad n=0,1;\quad w = -\Gamma  + 2\pi. $$

\noindent
From the energy profile

\begin{eqnarray}\label{34}
e(w,\Gamma ) = \frac{E}{E_s} = 1 - \cos \frac{w-\Gamma }{2}
+ \frac{h}{2} \sin^2 \frac{w}{2},
\end{eqnarray}

\noindent
and the matching condition (\ref{30}) one finds

\begin{eqnarray}\label{35}
\frac{d e(w(\Gamma),\Gamma)} {d\Gamma}= \frac{h}{4} \sin w =
-\frac{1}{2}\sin\frac{w-\Gamma }{2}.
\end{eqnarray}

\noindent
The inflection points for $h=2$ are $(\Gamma _c = 3\pi/2 , w_c =
\pi/2)$, $(\Gamma _c = \pi/2 , w_c = 3\pi/2)$
and a corresponding ground state is $(\Gamma_c =3\pi/2, w_B =
11\pi/6)$. Thus one finds

\begin{eqnarray}\label{36}
 F(h = 2-0)  = \frac{Q E_s}{2\pi } \left[ e(w_c,
\Gamma _c) - e (w_B, \Gamma _c) \right] = \frac{Q E_s}{4\pi } \cos
\frac{\pi }{6}=\frac{3\sqrt3}{8\pi} Q E_s.
\end{eqnarray}

\noindent
For $h = 2+0$ the path to follow in the integral

\begin{eqnarray}\label{37}
F  =- \frac{Q E_s}{2 \pi} \int^{2\pi}_0 d \Gamma \,\,
\frac{de}{d\Gamma} = -\frac{QE_s}{8\pi} h
\int^{2\pi}_0 d\Gamma  \sin w(\Gamma)
\end{eqnarray}

\noindent
is continuous and goes from $w(0) =2\pi$ to $w(\Gamma _c) = 3\pi/2$ and
then back to $w(2\pi) =2\pi$ resulting in a phase change of $\Delta
\varphi_i = -2 \pi$. The integration gives

\begin{eqnarray}\label{38}
F (h = 2+0) = \frac{Q E_s}{\pi}.
\end{eqnarray}

\noindent
This is the value of the force for all $h>2$ as pointed out in \cite{L94}.

Further analytical results are found for $h<2$. From the topological
condition $d\Gamma/dw=0$ of the critical inflection point  one gets

\begin{eqnarray}\label{40}
e(w_c,\Gamma_c) = 1+\frac{h}{4}-\frac{3}{4} \sqrt{\frac{4-h}{4}},
\end{eqnarray}

\noindent
with

\begin{eqnarray}\label{42}
w_c=\arccos\sqrt{\frac{4-h}{3h^2}}, \quad \Gamma_c=\arccos\sqrt{\frac{4-h}{3h^2}}
+2 \arccos\sqrt{\frac{n^2-1}{3}}.
\end{eqnarray}

\noindent
The perturbative expansion of the force $F(h)$ slightly below
$h=2$ then leads to

\begin{eqnarray}\label{44}
F (h) =F (h = 2-0) -\frac{QE_s}{4\pi}
\left\{\sqrt{3}+\frac{1}{2}\sqrt{\frac{1}{3}}\right\}\sqrt{2-h}+O(2-h).
\end{eqnarray}

\noindent
It is clear that the threshold for pinning is $h=1$ because there are no
metastable states for $h<1$.
The intermediate regime $1<h<2$ is treated numerically and the result
is shown in Fig. 3. The behaviour near $h
= 1$ is again $ F(h)  \sim (h - 1)^2 = \epsilon ^2$.
The critical exponent is thus $\beta = 2$ which seems to be universal for
local pinning.

\section{Discussion}

From its very definition local impurity pinning is expected to be
independent of spatial dimensionality $d$. This is born out by our
approach. Repeating the calculations of Sec. 2 for $d=2$ and $d=1$
always leads to the central equation (\ref{11}). However, the pinning
parameter $h \equiv h_d$ is not any more given by (\ref{7}) which refers
to $d=3$. Using half the the mean distance $\ell_d \gg \xi$ between
impurities as the distance from the impurity where $\varphi$ vanishes
it is found that $h_2=V_0 a_t \ln(\ell_2/\xi)/(\hbar v_t)$ for $d=2$.
For $d=1$ $h_1= \pi V_0 \ell_1/(2 \hbar v_F)$ is obtained.
The relation (\ref{22}) between the pinning parameter $h_d$ and
$\epsilon_i$ becomes $\epsilon_i = 2 \pi h_2 /\ln(\ell_2/\xi)$
for $d=2$ and $\epsilon_i= 4\,h_1$ for $d=1$.
These relations are similar to (\ref{22}) and do not change the conclusions
significantly. The case $d=1$ which does not require the short distance
cut--off $\xi$ is unrealistic since real CDW are never one--dimensional.

In summary we have studied a solvable model of local impurity phase pinning
which realizes the local pinning scenario in [5], namely singular
points, threshold behaviour,
and metastable states. In contrast to pinning in type II superconductors
static descreening and the possible deactivation of local pins at low
temperatures are unique to semiconducting CDW (and spin density) systems.
These results have been obtained within a phase only model plus some
amendments for descreening. Especially at low temperatures more general
models, e.g., those in \cite{A97} which take nonlinear screening (band
bending) into account may be considered.

\vspace{1cm}
\noindent
{\footnotesize{The authors thank P.B. Littlewood who initiated this study. They also thank
S.N. Artemenko for helpful discussions.}}

\newpage
\begin{figure}[ht]
\epsfxsize=240pt
\epsffile{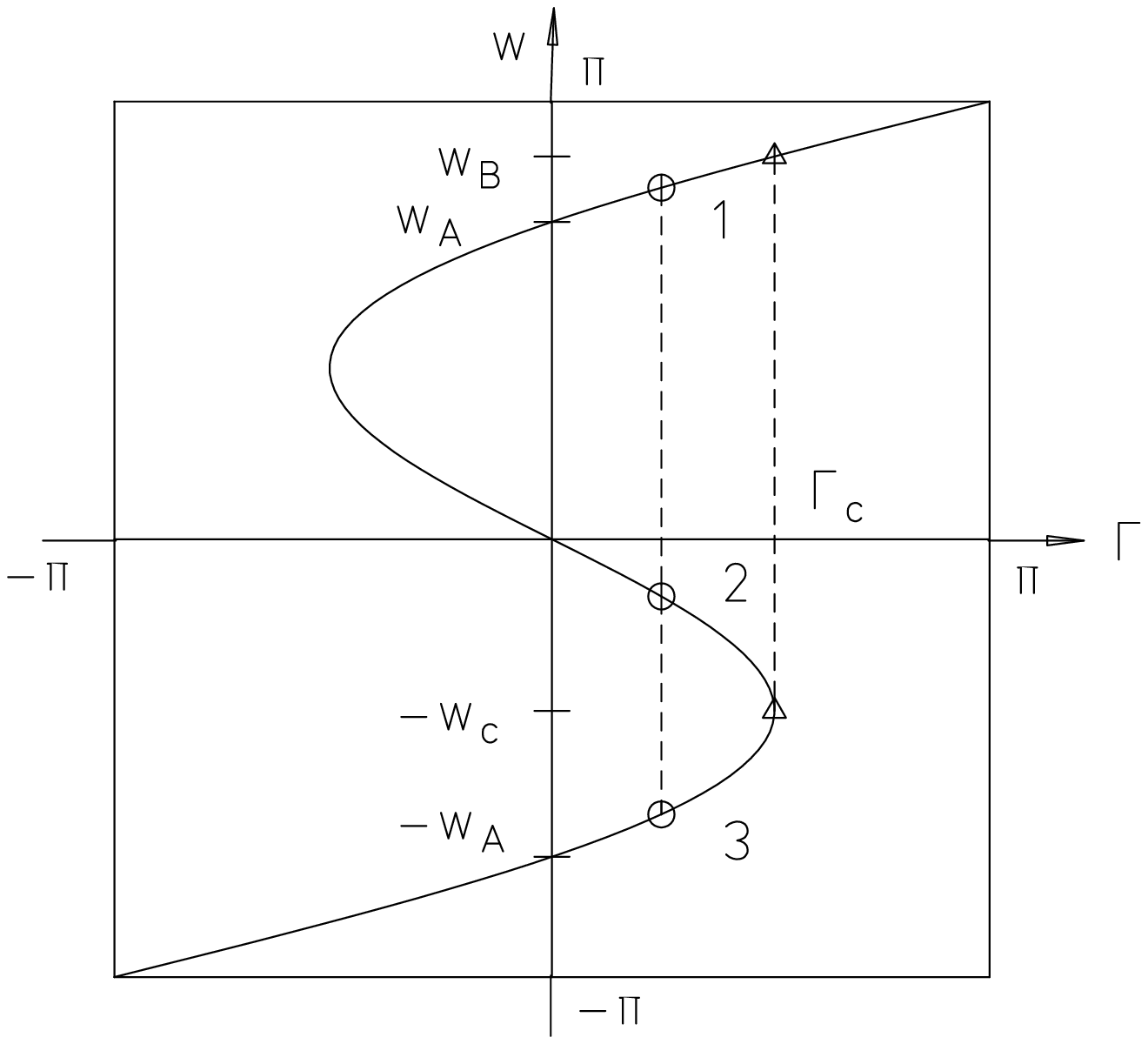}
\caption{Display of the solutions of the self consistency equation
(\ref{11}) in the form $w=\Gamma +h \sin w $ for $h=3$, i.e., not too far above
the pinning threshold $h=1$. The circles $1$, $2$, $3$ indicate the solutions
for $-\Gamma_c < \Gamma \le \Gamma_c$.
$(\Gamma_c,-w_c)$ is bifurcation point. The
pinning force results from the vertical transition from $-w_c$ to $w_B$ on the
upper branch.}
\end{figure}

\begin{figure}[ht]
\epsfxsize=240pt
\epsffile{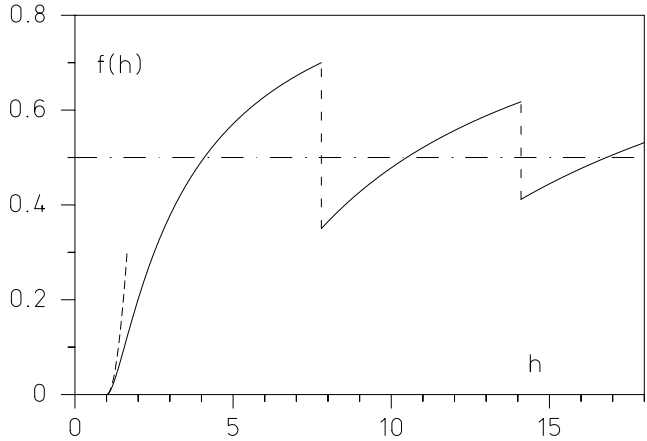}
\caption{Reduced pinning force $f$ in units of $QV_0$ as function of
pinning parameter $h$.
The local impurity phase pinning force $f$ sets in at $h=1$ with  exponent
$\beta=2$. Saturation value $f(h \rightarrow \infty)=1/2$,
the quadratic treshold behaviour, and the jumps at the points $h_n \approx
2 \pi n + \pi/2, n=1,2,...$ are indicated by broken lines.}
\end{figure}

\begin{figure}[ht]
\epsfxsize=240pt
\epsffile{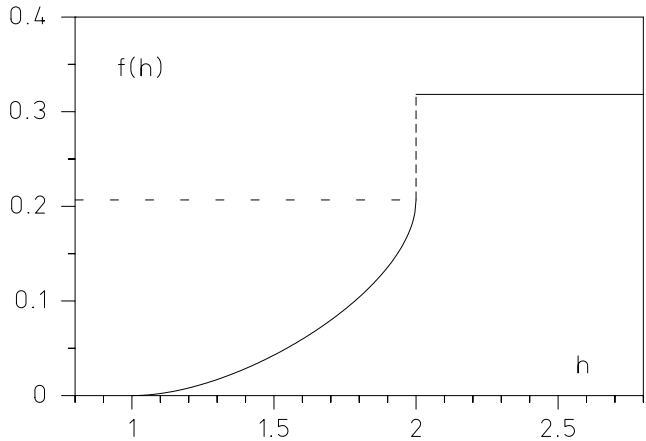}
\caption{Reduced pinning force $f(h)$ in units of $QE_s$ as function of
pinning
parameter $h$ for Larkin's model. There is no pinning force below $h=1$
and $f(h)$ jumps from $3\sqrt{3}/(8 \pi)$ to the final value $1/\pi$ at
$h=2$. Note the different normalization of $f(h)$ in comparison to Fig. 2.}
\end{figure}

\end{document}